\documentclass[twocolumn,showpacs,preprintnumbers,amsmath,amssymb]{revtex4}

\usepackage{graphicx}
\usepackage{dcolumn}
\usepackage{bm}
\usepackage{color}

\begin{document}

\preprint{YITP-09-30}

\title{Probing the birth of fast rotating magnetars through high-energy neutrinos}

\author{Kohta Murase$^{1}$, Peter M\'esz\'aros$^{2}$, and Bing Zhang$^{3}$}
\affiliation{%
$^{1}$Yukawa Institute for Theoretical Physics, Kyoto University,
Kyoto, 606-8502, Japan\\
$^{2}$Department of Astronomy \& Astrophysics; Department of Physics; Center for Particle Astrophysics;
Pennsylvania State University, University Park, PA 16802, USA\\
$^{3}$Department of Physics and Astronomy, University of Nevada at Las Vegas, Las Vegas, NV 89154, USA
}%

\date{March 27}
\begin{abstract} 
We investigate the high-energy neutrino emission expected from 
newly born magnetars surrounded by their stellar ejecta.  
Protons might be accelerated up to 0.1-100 EeV energies possibly 
by, e.g., the wave dissipation in the winds, leading to hadronic interactions 
in the stellar ejecta. 
The resulting PeV-EeV neutrinos can be detected by IceCube/KM3Net 
with a typical peak time scale of a few days after the birth of 
magnetars, making the characteristic soft-hard-soft behavior. 
Detections would be important as a clue to the formation mechanism 
of magnetars, 
although there are ambiguities coming from uncertainties 
of several parameters such as velocity of the ejecta.   
Non-detections would also lead to useful constraints on the scenario.
\end{abstract}

\maketitle
\section{\label{sec:level1}Introduction}
Magnetars are neutron stars endowed with the strongest magnetic fields 
known in the universe, $B \sim {10}^{14-15}$ G
(for reviews see \cite{WT04}). 
It is believed that 
$\sim 10$ \% of young galactic neutron stars possess such strong 
fields (for reviews see \cite{WT04}).  
Although the precise origin of these strong fields is uncertain, 
It has been argued that their amplification occurs 
via the dynamo mechanism during the Kelvin-Helmholtz cooling 
time, $t_{\rm KH} \sim 10-100$ s, in the protoneutron star 
(PNS) phase \cite{DT92}. The dynamo efficiency is 
partly determined by the initial rotation rate of 
$\Omega_{i} = 2 \pi/P_{i}$, and analytical estimates suggest 
that the formation of global strong magnetic fields might 
require $P_{i} \sim 1$ ms at birth. This will affect the dynamics 
of the PNS wind, by providing a significant reservoir of 
rotational energy comparable to that of the accompanying
supernova (SN) explosion. The strong magnetic fields can 
facilitate the SN explosion, and may be responsible for the
more powerful sub-class of SNe known as hypernovae (HNe), 
related to long gamma-ray bursts (GRBs) (e.g. \cite{Mes06} 
for reviews), which together with the PNS wind acts as a piston 
on the compressed stellar ejecta, i.e. the young expanding SN 
remnant (SNR) \cite{TCQ04}.
 
Newly born magnetars with rapid rotation rates may be
efficient ultra-high-energy cosmic ray (UHECR) accelerators
\cite{Aro03,Ghi+08}. Particle acceleration in neutron stars 
has been considered  
at both the polar or outer gap inside the magnetosphere, 
and near the wind zone \cite{WT04,Bed02}.
If ions are accelerated as well as electrons,
up to sufficiently high energies, 
they can produce high-energy neutrinos via, e.g., the $p\gamma$
or $pp$ reaction. Based on the polar gap models, Refs. 
\cite{Zha+03} discussed neutrino production 
via the $p\gamma$ reaction between ions and 
surface x rays, or by curvature pion radiation of ions.  
In this work, we investigate the high-energy 
magnetar neutrino emission resulting from a 
different scenario suggested by Ref. \cite{Aro03}, 
where cosmic-ray acceleration is attributed to 
the wake-field acceleration mechanism beyond 
the light cylinder. We consider the 
$pp$ and $p \gamma$ interactions of cosmic-ray ions with cold SNR 
nucleons and thermal photons. In the case of fast rotating magnetars, 
we find that the resulting neutrino energy fluence peaks around 
$t \sim \rm{a~few~days}$ due to hadronic cooling of mesons and muons. 
Such high-energy neutrinos can be detected by 
future $\rm{km}^{3}$ telescopes such as IceCube/KM3Net \cite{Ahr+04} 
in a few years or if magnetars are born at $\lesssim 10$ Mpc.    

\section{\label{sec:level2}The Model}
In the dynamo scenario, newly born magnetars have a large 
rotational energy \cite{DT92}. 
During $t_{\rm KH}$ the PNS winds would be thermally neutrino-driven or 
magnetically dominated but subrelativistic, and the spin 
down rate may be enhanced by neutrino-driven mass loss \cite{TCQ04}. 
After $t_{\rm KH}$, the winds become magnetically dominated and
relativistic, similarly to the case of pulsar winds. The rotational 
energy is extracted by the Poynting flux and gravitational waves \cite{Aro03}.
Using the vacuum dipole formula, the rotational energy loss rate by the magnetic 
wind at $t (> T_{\rm EM} \simeq {10}^{2.5}~{\rm s}
~I_{45}
~\mu_{33}^{-2}~{\Omega}_{i,4}^{-2})$ in the stellar frame can be estimated as
$L(t) \simeq 6.1 \times {10}^{47}~{\rm erg}~{\rm s}^{-1}
~{I}_{45}^2
~{\mu}_{33}^{-2}~t_4^{-2}$,  
where $\mu \equiv \frac{1}{2} B_{\rm NS} R_{\rm NS}^3 \simeq 0.5 \times {10}^{33}~{\rm G}
~{\rm cm}^{3}~B_{\rm NS,15} R_{\rm NS,6}^3$ is the magnetic dipole moment and 
the moment of inertia $I$ is set to ${10}^{45}~{\rm g}~{\rm cm}^2$
throughout this work.
The particles are expected to gain a fraction of the magnetar rotational energy 
during their acceleration, 
by tapping a fraction of the open field line voltage  
on their ways from the magnetar to the outside region.
The maximum cosmic-ray energy accelerated at $t (> T_{\rm EM})$ is 
\begin{equation}
E^{M}(t) = \eta Z e \Phi_{\rm mag} \simeq 2.0 \times {10}^{20}~{\rm eV}~Z~\eta_{-1}
~I_{45}
~\mu_{33}^{-1} t_4^{-1} \label{Emax},
\end{equation}
where $\eta$ parameterizes the uncertainties in the utilization of 
the potential drop. 
In polar gap models, the parallel electric fields would be 
significantly screened in very young pulsars, implying 
$\eta \ll 0.1$ \cite{WT04,Zha+03}. Nevertheless, as in pulsars, 
a significant fraction of the Poynting energy could be converted to 
the kinetic energy well outside the light cylinder, via 
mechanisms such as surf-riding acceleration. Ref. \cite{Aro03} 
suggested UHECR acceleration by the wake-field acceleration 
in the equatorial wind, where it was argued that, 
for an oblique rotator, much of 
the Poynting flux might be tied up in waves and the crinkled 
frozen-in current sheets dissipate around $r_{\rm diss} 
\sim {10}^{3-4} (c/\Omega)$. 
If wave emission is the relevant dissipation process, 
it might form large amplitude electromagnetic waves pushing 
ions by the ponderomotive force, 
$F_{\rm pond} \approx mc \Omega (Z e \delta B/m c \Omega)$ 
\cite{CTT02}.
Hence, the work on a particle moving a distance 
$l$ through the wave is $F_{\rm pond} l \approx 
\eta Z e \Phi_{\rm mag}$, as long as 
$\delta B \sim B$ and $\eta \equiv l/r \sim 0.1$. 
Following Ref. \cite{Aro03}, we hereafter 
assume that such prompt cosmic-ray acceleration mechanism
is in operation, and use Eq. (\ref{Emax}). 
The exact nature of the dissipation and acceleration mechanisms 
in the wind 
is currently uncertain and a detailed study 
is beyond the scope of this paper. Approximately, 
since we may expect that the ion injection 
rate around the equatorial sector is 
the Goldreich-Julian rate \cite{GA94}, 
the cosmic-ray spectrum can be written as \cite{Aro03,BEO00}
\begin{equation}
\frac{d N}{d E} = \frac{9}{8} \frac{c^2 I}{Z e \mu}
\frac{1}{E(1+E/E_G)},
\label{spectrum}
\end{equation}
where $E_G \equiv (5/72)(\eta Z e \mu^3/G I^2 \epsilon^2)$ 
and, for simplicity, we have assumed 
that all 
the particles accelerated at $t$ have $E^{M}(t)$.  
Note that since the acceleration is expected to occur 
promptly, energy losses can be neglected. Hence
the adiabatic and radiation losses in the wind 
become irrelevant since accelerated ions will not
be coupled to the fields and their curvature radius 
is large enough \cite{Aro03}. 


\section{\label{sec:level4}The Neutrino Spectrum and Flux}
A newly born magnetar will be surrounded by the young SNR, separated
by a cavity evacuated by the wind or rapidly expanding SN shock.
The particle acceleration in the wind occurs around $r_{\rm acc} 
\sim r_{\rm diss} \sim {10}^{10.5}~{\rm cm}~{\mu}_{33} {t}_{4}^{1/2}$
(hereafter we consider the cosmic-ray ions to be protons \cite{GA94}).
The wind termination shock and the SN shock radii are both larger 
than $r_{\rm acc}$ for sufficiently late times as considered here. 
Thus, a significant fraction of cosmic rays will 
interact with the stellar ejecta, unless the latter 
is punctured or disrupted by the wind itself (see below for general
cases). First, we consider the interaction between cosmic rays and SNR nucleons
assuming that cosmic rays are emitted isotropically, to obtain conservative results.
 
The cosmic rays will interact with the SNR via $pp$ reactions, producing mesons. 
When the (magnetar-powered) SN shock has a high velocity of 
$\beta_{\rm SN} c$, the effective optical depth for the $pp$ reaction is 
$f_{pp} \approx {\kappa}_{pp} \sigma_{pp} n_p \Delta_{\rm SN} \simeq 5.7 \times {10}^{4}~
M_{\rm SN,1} \beta_{\rm SN,-1}^{-2} t_{4}^{-2}$, 
where $\kappa_{pp} \sim 0.5-0.6$, $\sigma_{pp} \sim {10}^{-25}~\rm{cm}^{-2}$ 
(in the 100 PeV range) \cite{Fle+94}, and $\Delta_{\rm SN} 
\sim r_{\rm SN} \approx \beta_{\rm SN} c t 
\simeq {10}^{13.5}~{\rm cm}~\beta_{\rm SN,-1} t_{4}$. 
The meson production efficiency is estimated as 
$f_{\rm mes} \sim {\rm{min}}(1,f_{pp})$.
Since the effects of magnetic fields in the SNR can typically be neglected, 
the interaction time is $t_{\rm int} \approx {\Delta}_{\rm SN}/c$.
 
The resulting charged mesons decay into neutrinos via $\pi^{\pm}
\rightarrow e^{\pm}+{\nu}_{e}({\bar{\nu}}_{e})+{\nu}_{\mu}+{\bar{\nu}}_{\mu}$.
The neutrino spectrum roughly follows the cosmic-ray spectrum,
but the high-energy spectrum is modified when mesons and muons 
cool down before they decay. In our cases, the inelastic 
$\pi p/\mu p$ collision can be relevant. When $f_{\pi p} 
\approx t_{\rm int}/t_{\pi p} \gtrsim 1$, the pion neutrino 
flux is suppressed as $f_{\rm sup} \sim {\rm min} (1, 
(t_{\pi p}/\gamma_{\pi} \tau_{\pi}))$, where 
the break energy is $E_{\nu}^{\rm had}~\approx 
0.25~(t_{\pi p}/\tau_{\pi}) m_{\pi} c^2 
\simeq 32~{\rm TeV}~M_{\rm SN,1}^{-1} \beta_{\rm SN,-1}^{3} t_{4}^{3}$. 
$\tau_{\pi/\mu}$ is the proper life time of charged pions/muons 
and $t_{\pi p/\mu p} \approx {(\kappa_{\pi p/\mu p} 
\sigma_{\pi p/\mu p} n_p c)}^{-1}$ is their hadronic cooling time. 
Since the neutrino spectrum is proportional to 
$f_{\rm mes} f_{\rm sup}$, 
its rough expression for $dN/dE \propto E^{-p}$ is 
\begin{eqnarray}
E_{\nu}^2 \phi_{\nu} 
\propto \left\{ \begin{array}{ll}
{(E_{\nu}/E_{\nu}^{\rm had})}^{2-p} 
& \mbox{(for $E_{\nu} \leq E_{\nu}^{\rm had}$)} \\
{(E_{\nu}/E_{\nu}^{\rm had})}^{1-p} 
& \mbox{(for $E_{\nu}^{\rm had}< E_{\nu} \lesssim \frac{1}{4} E^{\rm max}$)}
\end{array} \right. \label{one}
\end{eqnarray}
The resulting neutrinos can propagate in the SNR matter without 
significant attenuation because 
$\tau_{\nu p} \approx \sigma_{\nu p} n_p \Delta_{\rm SN} 
\simeq {0.017}~E_{\nu, \rm EeV}^{0.363} 
M_{\rm SN,1} \beta_{\rm SN,-1}^{-2} t_{4}^{-2} \ll 1$. 

\begin{figure}[bt]
\includegraphics[width=\linewidth]{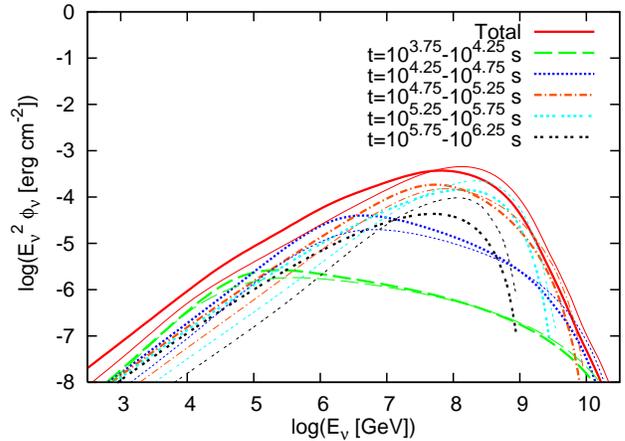}
\caption{\small{\label{Fig1} 
The $(\nu _{\mu} + \bar{\nu} _{\mu})$ fluence from a newly born magnetar 
at $5$ Mpc, at different time intervals. The fluence peaks around 
$t \sim 2$ days because hadronic cooling of mesons and muons is important 
at earlier times, while the amount of 
cosmic rays decreases with time. 
Thick/thin lines are for the cases without/with the radiation field. 
Vacuum neutrino oscillations are considered.
}}
\end{figure}

We have performed detailed numerical calculations to
evaluate neutrino spectra and fluxes through the method of 
Refs. \cite{Mur07}. The calculations are performed during 
a time $t_{\rm int}$,
taking into account the high meson multiplicity of the 
high-energy $pp$ reaction, based on the SYBILL code 
\cite{Fle+94}, and hadronic cooling of 
pions and muons with the approximated cross sections
of $\sigma_{\pi p} \simeq 5 \times {10}^{-26}~{\rm cm}^2$ 
and $\sigma_{\mu p} \simeq 2 \times {10}^{-28}~{\rm cm}^2$ \cite{RMW03}.
We neglect contributions from 
meson production via $\pi p/\mu p$ processes which can affect 
the spectra by factors $\mathcal{O}(1)$, since their influence 
is modest and only at relatively early times, $t \lesssim \rm day$. 
As shown later, the detailed spectra will be somewhat 
different from Eq. (\ref{one}) especially at high energies, due to 
the high meson-multiplicity of the $pp$ reaction and the accumulation of 
cooled mesons and muons.
The only necessary input quantities are the cosmic-ray flux
and target nucleon density. 
The former is given by Eqs. (\ref{Emax}) 
and (\ref{spectrum}) and the latter by $M_{\rm SN}$ and $\beta_{\rm SN}$. 

In Figs. 1 and 2, the resulting spectra and light curves are shown for 
$M_{\rm SN}=M_{\rm SN,1} \equiv 10 M_{\odot}$ and $\beta_{\rm SN}=0.1$. 
We can see that the neutrino energy fluence becomes maximal
around $t \sim 2~{\rm days}$, because of hadronic cooling of pions and 
muons at earlier times and a decrease of $E^2 \frac{dN}{dE}(t) \propto 
E^M/\mu \propto t^{-1}$ at later times. 
As a result, we expect a hardening spectrum at $t \lesssim 
2~\rm days$, and a softening spectrum at $t \gtrsim 2~\rm days$ 
in the $100~\rm TeV - 10~\rm PeV$ range (Fig. 2). 
The peak time $T_{\rm had}$ is determined by equating $E_{\nu}^{\rm had}$ 
with the typical neutrino energy $E_{\nu}^{\rm typ} 
\sim 0.03 E^M$ \cite{Fle+94}. We have 
$T_{\rm had} \sim 2 \times {10}^{5}~{\rm s}~{\eta}_{-1}^{1/4}
\mu_{33}^{-1/4} {\beta}_{\rm SN,-1}^{-3/4} M_{\rm SN,1}^{1/4}$ 
and the corresponding peak energy 
$E_{\nu,\rm pk}^{\rm had} \sim 300~{\rm PeV}~{\eta}_{-1}^{3/4}
\mu_{33}^{-3/4} {\beta}_{\rm SN,-1}^{3/4}
M_{\rm SN,1}^{-1/4}$ which agree with Figs. 1 and 2. 
Note that cases of $T_{\rm had} \gg T_{\rm EM}$ are considered. 
The main contribution comes from the cosmic rays produced
at $t \sim T_{\rm had}$. Since $E_{\nu}^2 \phi_{\nu} (t) 
\propto E^2 \frac{dN}{dE} (t) \propto E^M/\mu$, the neutrino fluence 
per flavor around the peak time is roughly estimated as
$\sim {10}^{-4}~{\rm erg}~{\rm cm}^{-2}~D_{5~{\rm Mpc}}^{-2}
f_{\rm  mes} f_{\rm sup} {\eta}_{-1}^{3/4}
\mu_{33}^{-7/4} {\beta}_{\rm SN,-1}^{3/4}
M_{\rm SN,1}^{-1/4}$. The total expected 
muon event rates (above ${10}^{0.5}$ TeV)
by IceCube is $N_{\mu} 
\sim 2~D_{5~{\rm Mpc}}^{-2}$ events in two days, which 
will be more than the atmospheric neutrino-induced event 
rates within $1^{\circ}$, $N_{\mu}^{\rm atm} 
\sim {10}^{-2.5}$ events/day. Magnetars arising
at distances closer than 5 Mpc would yield higher fluxes
observed as neutrino multiplets, which allow us to recognize them 
as signals without coincident detections with photons 
and even to see the characteristic soft-hard-soft behavior.
Since the magnetar birth rate is $\sim 
{10}^{-3}~{\rm yr}^{-1}~{\rm galaxy}^{-1}$, the probability 
to encounter a birth 
is non-negligible. From the number of local galaxies, 
we expect $\sim 0.02-0.05~{\rm yr}^{-1}$ for the birth of 
magnetars within 5 Mpc \cite{ABY05}. 

One may expect an additional radiation field, leading to 
$p \gamma$ neutrinos in addition to $pp$ neutrinos. For example, 
if the magnetar wind drives the SN explosion in its birth 
\cite{TCQ04}, a significant fraction of the outflow energy 
may be dissipated as radiation via the shocks. 
(The radiation field can also be expected in case of GRB jets 
in the star \cite{MW01}.) 
Therefore, we also show the case where the radiation field is included.  
In Figs. 1 and 2, the case for $k T_{\gamma} \simeq 
0.4~{\rm keV} \epsilon_{\gamma}^{1/4} E_{\rm exp, 51}^{1/4} 
\beta_{\rm SN, -1}^{-3/4} t_{3}^{-3/4}$ is also shown. 
Here $E_{\rm exp}$ is the outflow energy and $\epsilon_{\gamma}$ 
is the radiation efficiency. When the radiation 
field exists in the SN ejecta, the previous expression of 
$f_{\rm mes}$ should be replaced with 
$f_{\rm mes} \sim {\rm{min}}(1, {\rm max}(f_{pp}, f_{p \gamma}))$, 
where the effective optical depth for the photomeson production, 
$f_{p \gamma}$, is roughly estimated as 
$f_{p \gamma} \approx {\kappa}_{p \gamma} \sigma_{p \gamma} 
n_\gamma \Delta_{\rm SN} \simeq 380~{\epsilon}_{\gamma}^{3/4} 
E_{\rm exp,51}^{3/4} \beta_{\rm SN,-1}^{-5/4} t_4^{-5/4}$
around the $\Delta$-resonance energy of 
$E_{\Delta} \simeq 2.4~{\rm PeV}~{\epsilon}_{\gamma}^{-1/4} 
E_{\rm exp,51}^{-1/4} \beta_{\rm SN,-1}^{3/4} t_4^{3/4}$. 
Here,  $\kappa_{p \gamma} \sim 0.2$, $\sigma_{p \gamma} 
\sim 5 \times {10}^{-28}~\rm{cm}^{-2}$ at the $\Delta$-resonance.
Correspondingly, the expression of $f_{\rm sup}$ includes 
the cooling of mesons and muons due to interactions with photons
as well as their hadronic cooling.  
Following Ref. \cite{Mur07}, neutrino spectra are numerically 
calculated, taking into account the radiation field.
Although the radiation field can change spectra 
as a result of the difference in the meson multiplicity, 
we may expect that the total energy fluence around the peak 
energy and the qualitative feature are similar.  

\begin{figure}[tb]
\includegraphics[width=\linewidth]{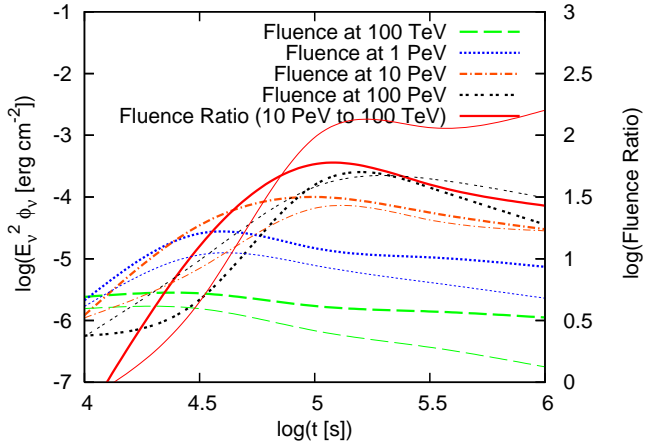}
\caption{\small{\label{Fig2}
Neutrino lightcurves corresponding to Fig. 1 at
100 TeV (dashed line), 1 PeV (dotted line), 10 PeV (dotted-dashed line), 
100 PeV (double dotted line), and the ratio of 10 PeV fluence to 
100 TeV fluence (thick line). Cooling of mesons and 
muons is important at $t \lesssim 2$ days, while the amount of 
accelerated protons decreases with time. 
Thick/thin lines are for the cases without/with the radiation field. 
}}
\end{figure}

Next, let us consider the sum of neutrinos from individual 
magnetars, i.e., the \textit{cumulative} neutrino background. 
The typical magnetar rate would be $\sim 10$ \% of core-collapse 
(CC) SN rate, $R_{\rm SN}(0) \sim 1.2 \times {10}^{5}~{\rm Gpc}^{-3}
~{\rm yr}^{-1}$ \cite{WT04,MDP98}. Possibly, the birth rate of fast 
rotating magnetars may be comparable to that of HNe that 
may be powered by magnetars, implying $R_{\rm HN}(0) \sim 2 
\times {10}^{3}~{\rm Gpc}^{-3}~{\rm yr}^{-1}$ \cite{MDP98}.  
By using our numerical results, the cumulated fluxes can be 
estimated as \cite{Mur07,WB97}
\begin{eqnarray}
E_{\nu}^2 \Phi_{\nu} &\sim& 
3 \times 10^{-9}~{\rm{GeV cm^{-2}~s^{-1}~str^{-1}}}
f_{\rm{mes}} f_{\rm{sup}}
\eta_{-1}^{\frac{3}{4}} \mu_{33}^{-\frac{7}{4}} 
\nonumber\\
&\times& 
\beta_{\rm SN,-1}^{\frac{3}{4}} M_{\rm SN,1}^{-\frac{1}{4}}
\frac{f_{\rm geo}}{0.5} \frac{f_{z}}{3}
\frac{R_{\rm{mag}}(0)}{1.2 \times {10}^{3}~\rm{Gpc}^{-3}
~\rm{yr}^{-1}},
\end{eqnarray}
where $f_{\rm geo}$ is the fraction of the magnetars with the preferred 
geometry for ion acceleration \cite{Aro03} and $f_{z}$ expresses the 
contribution from the high redshift sources \cite{Mur07,WB97}. 

The numerical results are shown in Fig. 3, where the birth rate 
evolution is included with the SFR2 model for 
magnetars \cite{Mur07}. The muon event rates are $N_{\mu} \sim 18$ 
events/yr for $R_{\rm mag}=0.1 R_{\rm SN}$ and $N_{\mu} \sim 4$ events/yr 
for $R_{\rm mag}=R_{\rm HN}$, respectively.
Note that cross-correlation studies between neutrinos 
and CC SNe/HNe can be important, but we may expect one 
time- and space-coincident event among $\sim {10}^{4-5}$ magnetar births. 
However, from Fig. 3, we still expect future constraints on
the ``diffuse" neutrinos are important to test the magnetar 
scenario 
since expected neutrinos have rather high energies. 

\begin{figure}[tb]
\includegraphics[width=\linewidth]{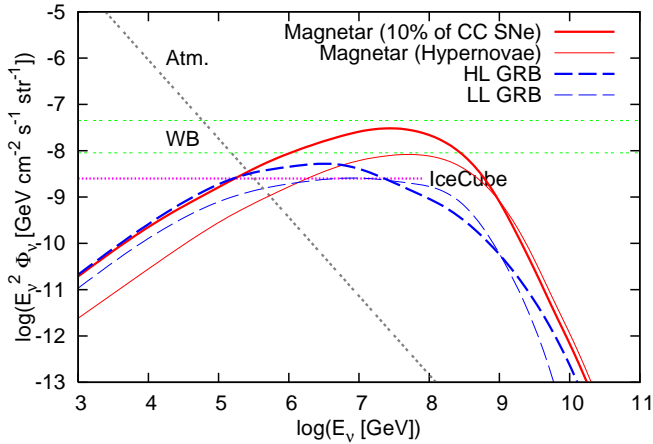}
\caption{\small{\label{Fig3} 
The cumulative $(\nu _{e} + \bar{\nu} _{e}+
\nu _{\mu} + \bar{\nu} _{\mu}+\nu _{\tau} + \bar{\nu} _{\tau})$ 
background from newly born magnetars. 
Thick solid line: Magnetar (10\% of CC SNe) shows the case where 
10\% of CC SNe bear fast rotating magnetars, 
with $M_{\rm SN}=10M_{\odot}$. 
Thin solid line: Magnetar (Hypernovae) is a case where 
the fast rotating magnetar rate is comparable to the HN rate,
with $M_{\rm SN}=M_{\odot}$.
For comparison, various GRB neutrino backgrounds for the GRB-UHECR hypothesis 
are also shown. 
Thick dashed line: HL GRB, prompt neutrinos from HL GRBs \cite{Mur07,WB97}.
Thin dashed line: LL GRB, prompt neutrinos from LL GRBs \cite{Mur07}.
WB: The Waxman-Bahcall bounds are shown as benchmarks \cite{WB97}.
Atm: the conventional atmospheric neutrino background.
A $\Lambda$CDM cosmology ($\Omega _{\rm{m}}=0.3, 
\Omega _{\Lambda}=0.7; H_{0}=71 \, \rm{km\, s^{-1} \, Mpc^{-1}}$)
and $z_{\rm{max}}=11$ is used.
}}
\end{figure}

We consider the possible beaming effects of 
the cosmic rays, SNR puncture and disruption by the winds, 
since these can affect the escape and detectability 
of the cosmic rays.
Cosmic rays themselves may be beamed, enhancing the
neutrino signals from individual sources by
the beaming factor while the background is unchanged. 
In addition, if the PNS winds are significantly collimated, 
they may puncture the stellar envelope, leading to long 
GRB jets \cite{TCQ04}. 
Only cosmic rays 
that are emitted along the penetrating jets 
can escape without depletion. However, even if
cosmic rays are beamed along the jet, we still 
expect neutrino production 
when jets are choked 
rather than successful \cite{MW01}.  

The disruption resulting in the formation of supershells 
expanding into the interstellar medium 
was discussed in Ref. \cite{Aro03}. 
Such phenomena have never been observed in CC SNe/HNe, 
but we discuss them for the sake of generality. 
The effect of the disruption by energy shedding may be 
characterized by a clumpiness factor, 
$C \equiv \delta \rho/\rho$. 
The no shedding case corresponds to $C=1$. 
In clumpy SNRs, neutrinos are produced when clumps 
lie along the line of sight. 
While values of $C$ are uncertain, as an example
$C \sim 3^3$ implies that the probability 
to see neutrinos is $\sim C^{-2/3} \sim 1/9$,  
and the background is similarly reduced. 

\section{\label{sec:level5}Implications and Discussions}
The birth of fast rotating magnetars is a common scenario 
discussed in connection with HNe and GRBs
\cite{DT92}, although the origin of their strong magnetic fields 
is controversial (e.g., dynamo vs fossil fields). Here we have 
shown that high-energy neutrino signals can serve as the smoking gun 
signal announcing the birth of fast rotating magnetars. 
A suppression of the highest-energy neutrinos at 
$t \lesssim T_{\rm had} \sim {\rm a~few~days}$ 
implies that cosmic-ray acceleration occurs inside the SNR, 
and a characteristic soft-hard-soft behavior is expected. 
Although our predicted fluxes are below the current observational limits, 
they can be tested by IceCube, KM3Net, ARIANNA and Auger in the near future. 
Even non-detections would provide useful constraints on 
this magnetar scenario.
 
There are also other possible ways to produce neutrinos. 
First, at earlier times ($\lesssim$ days), there may be radiation fields with 
the temperatures of $0.1-1$ keV (at $r_{\rm SN} \sim {10}^{13.5}$ cm), in the 
shocked stellar ejecta ahead of the wind \cite{MW01,RMW03} or in the cavity \cite{WLW93}. 
The former case is also demonstrated in this work.
Collimated winds launched at very early times ($\lesssim 10^3$ s)
may become successful jets such as GRB jets. 
Then, a fraction of the cosmic rays interact with 
late internally dissipated or external-shock photons, 
making other very high-energy neutrino signals, 
similarly to the case of GRBs \cite{Mur07}. 

Cosmic rays and neutrinos could be expected possibly also in 
\textit{normal} pulsars  
where $T_{\rm had} \ll T_{\rm EM}$ is anticipated. 
When they have weaker magnetic fields but rapid rotation speeds, 
the peak time of the energy fluence can be later than days, and 
then neutrino emission lasts longer. 
Since it was not the focus of this work, see Refs. \cite{Bed02} 
for comprehensive discussions.

Magnetar neutrinos may be useful for revealing the possible connection 
between magnetars and UHECRs. Since our purpose here is not an 
explanation of UHECRs, we make only two brief comments on this possibility: 
(a) only a fraction of magnetars can be UHECR sources, and (b) the 
Auger spectrum seems to conflict with Eq. (\ref{spectrum}) \cite{Abr+08}. 
Concerning point (a), one may think of many possible reasons 
such as initial rotation rate, geometry, puncture or disruption of the SNR. 
Interestingly, the rate required in the magnetar scenario is comparable to the 
HN rate \cite{Ghi+08,Abr+08}. 
Point (b) requires more careful consideration, but we may expect 
a realistic spectrum to differ from Eq. (\ref{spectrum}), 
depending on the detailed mechanism.  
For example, if test particles are stochastically accelerated by waves, 
$dN/dE \propto E^{-2}$ can be expected \cite{CTT02}.
Although this requires further investigations beyond 
the scope of this work, the use of Eq. (\ref{spectrum}) is 
sufficient for demonstrations. Other cases can easily be 
predicted, once the cosmic-ray spectrum is given.  


\begin{acknowledgements}
K.M. thanks K. Kohri for helpful discussions.
We acknowledge support by a Grant-in-Aid from JSPS, and by a Grant-in-Aid for 
the Global COE Program "The Next Generation of Physics, Spun from Universality 
and  Emergence" from MEXT (K.M.), by NASA NNX08AL40G, NSF PHY -0757155 (P.M.) 
and NASA NNG05GB67G (B.Z.).
\end{acknowledgements}

\appendix


\end{document}